\documentclass[aps,nofootinbib,prl,showpacs,onecolumn,groupedaddress,superscriptaddress]{revtex4}

\usepackage{graphicx}
\usepackage{dashrule}
\usepackage{dcolumn}
\usepackage{mathtools}
\usepackage{amssymb}
\usepackage{bbold}
\usepackage{amsmath}
\usepackage{amsfonts}
\usepackage{wasysym}
\usepackage{slashed}
\usepackage[dvipsnames]{xcolor}
\usepackage{soul}
\usepackage{rotating}
\usepackage{paracol}
\usepackage{nicefrac}
\usepackage{MnSymbol}
\usepackage{cleveref}

\newcommand{\figref}[1]{fig.~(\ref{#1})}

\newcommand{\eref}[1]{\cref{#1}}
\newcommand{\eg}{\textit{e.g.}~}
\newcommand{\ie}{\textit{i.e.}~}

\newcommand{\eftnopi}{\mbox{EFT($\slashed{\pi}$) }}
\newcommand{\ve}[1]{\ensuremath{\boldsymbol{#1}}}

\usepackage{xcolor}

\usepackage{soul}

\begin{document}

\title{Emergence of $^4$H $J^\pi=1^-$ resonance in contact theories}

\author{Lorenzo Contessi}
\address{Universit\'e Paris-Saclay, CNRS-IN2P3, IJCLab, 91405 Orsay, France}
\address{IRFU, CEA, Universit\'e Paris-Saclay, 91191 Gif-sur-Yvette, France}

\author{Martin Sch{\"a}fer}
\address{The Racah Institute of Physics, The Hebrew University, Jerusalem 9190401, Israel}

\author{Johannes Kirscher}
\address{Department of Physics, SRM University - AP, Amaravati 522502, Andhra Pradesh, India}
\address{Theoretical Physics Division, School of Physics and Astronomy,
  The University of Manchester, Manchester, M13 9PL, UK}
\address{Institute for Nuclear Studies, Department of Physics,
The George Washington University, Washington DC 20052, USA}

\author{Rimantas Lazauskas}
\address{IPHC, IN2P3-CNRS/Universit\'e de Strasbourg BP 28, F-67037 Strasbourg Cedex 2, France}

\author{Jaume Carbonell}
\address{Universit\'e Paris-Saclay, CNRS/IN2P3, IJCLab, 91405 Orsay, France}

\date{\today}

\begin{abstract}

We obtain the $s$- and $p$-wave low-energy scattering parameters 
for n$^3$H elastic scattering and the position of the $^4$H $J^\pi=1^-$ resonance using the pionless effective field
theory at leading order.
Results are extracted with three numerical techniques:
confining the system in a harmonic oscillator trap, solving the Faddeev-Yakubovsky 
equations in configuration space, and using an effective two-body
cluster approach. 
The renormalization of the theory for the
relevant amplitudes is assessed in a
cutoff-regulator range between $1\,\text{fm}^{-1}$ and
$10\,\text{fm}^{-1}$.

Most remarkably, we find a cutoff-stable/RG-invariant resonance in the $^4$H $J^\pi=1^-$ system.
This $p$-wave resonance is a universal consequence
of a shallow two-body state and the introduction of a three-body $s$-wave scale set by the triton binding energy.
The stabilization of a resonant state in a few-fermion system through pure contact interactions has a significant consequence for the powercounting of the pionless theory.
Specifically, it suggests the appearance of similar resonant states also in larger nuclei, like 16-oxygen, in which the theory's leading order does not predict stable states. 
Those resonances would provide a starting state to be moved to the correct physical position by the perturbative insertion of sub-leading orders,
possibly resolving the discrepancy between data and contact EFT.
\end{abstract}

\maketitle

\section{Introduction}

Contact effective field theory is a very successful approach in describing bosonic~\cite{Bazak:2016wxm,Carlson:2017txq,Bazak:2018qnu,Bedaque:1998kg,Bedaque:1998km} and few-nucleon~\cite{Bedaque:1999ve,Platter:2004zs,Platter:2004he,Barnea:2013uqa}~systems maintaining a tight connection to the underlying interaction~\cite{Hammer:2019poc,Barnea:2013uqa}. 
It promises a simple and clear framework to study the nuclear force. However, recent works suggest that the nuclear version, pionless effective field theory~\eftnopi, fails to predict stable bound states for $p$-shell nuclei\footnote{Fermionic systems with more particles than fermionic internal states} at the leading order (LO) \cite{Stetcu:2006ey,Contessi:2017rww,Dawkins:2019vcr,Schafer:2020ivj}. 
All these LO studies indicate
that beyond-$s$-shell nuclear systems,
experimentally observed as bound, clusterize 
into mutually unbound $s$-shell subcomponents.
For instance, one of the lightest bound $p$-shell
nuclei, $^6 {\rm Li}$, 
breaks into a $\rm ^4He$ in its ground 
state and a deuteron ($^2$H). In other words,
$^6 {\rm Li}$ is predicted to be unbound 
in the zero-range contact limit~\cite{Schafer:2020ivj}.
Nonetheless, it can be argued that this effect is
only due to the large LO truncation uncertainty,
and that the non-clustered state still exists as
a virtual state or a resonance above the threshold.
Only after the inclusion of higher order corrections
these poles may be brought to the bound-state region. 
However, to preserve~\eftnopi~renormalizability, 
this can only be done in perturbation theory.
Since perturbative insertions of higher order terms cannot
create any pole {\it ex novo}, the possible absence of such poles
in the contact limit and within convergence radius of the \eftnopi at
LO would put in question any chance of retrieving the correct nuclear
description beyond $s$-shell.
The question is thus simple: Does LO \eftnopi interaction
support the existence of such poles?

Historically, resonances were studied in (A$\le$4)-body bosonic systems close to the unitary limit \cite{Deltuva:2012ig,deltuva2020energies} confirming the existence of a tower of Efimov resonances. 
A recent study \cite{Habashi:2020qgw} proved the possibility of creating two-body $s$-shell resonances in the contact limit with negative effective range.
In LO \eftnopi~\\ framework, a resonance have been also found in neutral hypernuclear system $\Lambda$nn \cite{Schafer:2020rba} and $s$-wave virtual states have been confirmed as exited states of triton \cite{Rupak:2019} and hypertriton \cite{Schafer:2020rba}.
A recent study of $^3$n confirmed the absence of $p$-wave resonances or virtual states in this system \cite{Dietz:2021haj},
consistent with the absence of a finite scale that could fix the position of a resonance inside the convergence radius of the theory. 
For even larger $p$-shell nuclei, Deltuva et al. \cite{Deltuva:2011djw}
made a series of calculations of the four-body system employing a two-body gaussian potential of finite range, 
including the ($J^\pi=0^-$, $T=0$) four-nucleon state.
In this study multiple deep three-body bound states appear once reducing the range of the interaction and thus in the ultraviolet limit one has to choose as a reference three-body energy the one of the shallowest state. 
As long as there exist only one three-body state the resonance is clearly visible in scattering observables, while further reducing interaction range the effect of resonance  fades away. 
Therefore authors concluded that the resonance is due to finite cut-off effects.
Although in the last study position of the relevant $S$-matrix pole was not determined and the conclusion was drawn based on visual inspection of the calculated phase shifts. 
The major goal of this work is to study the emergent resonance states in $^4$H using a fully renormalizable theory as a function of the interaction cutoff.
The existence of an eventual cutoff-stable $p$-shell resonance using \eftnopi at LO would open the possibility of finding the same structure also in heavier nuclei, with the chance of stabilizing many-nucleon systems by perturbative sub-leading corrections.
Experimentally, $^4$H resonances are clearly seen in n$^3$H elastic cross sections \cite{PhysRev.119.1981,SEAGRAVE1972250}, $\pi^-$ absorption experiments \cite{Sennhauser:1981tr,Gornov:1991fg}, and transfer reactions \cite{Miljanic:1986zz,Sidorchuk:2003fwa}. 
However, the extracted resonance parameters differ substantially, depending on the used experimental protocol and the underlying hypothesis in the data analysis. 
For instance, results obtained by the same $\pi^-$+$^9$Be$\,\to\,$d+t+$^4$H experiment, performed by the same authors, yields for the lowest state ($E_R=3.0\pm0.2$~MeV,~$\Gamma=4.7\pm1$~MeV)~\cite{Gornov:1991fg}, ($E_R=2.0\pm0.2$~MeV,~$\Gamma=1.2\pm0.2$~MeV)~\cite{Gurov:2005sp},
and ($E_R=1.6\pm0.1$~MeV,~$\gamma^2=0.4\pm0.1$~MeV)~\cite{Gurov2005p}, where $\gamma$ is the reduced width (see the original paper), depending whether one assumes the experimental signal to be created respectively by one, two, or three resonant states.
Independently, the $R$-matrix analysis of n$^3$H cross section \cite{Tilley:1992zz} concluded about a superposition of several $p$-waves resonant states with quantum numbers $J^\pi$=$0^-$,$1^-_I$, $1^-_{II}$,$2^-$  \cite{Lazauskas:2004uq}. 
The groundstate parameters ($E_R$=3.2 MeV, $\Gamma$=5.4 MeV) are hardly compatible with the previous many-state analysis. 
The discrepancies among different experimental approaches further
increase when including the transfer reaction experiments.

From a numerical point of view, computing the $S$-matrix pole of a resonance is not an easy task. 
Microscopic calculations (in comparison with cluster calculations) for nuclear few-body systems are therefore not abundant and up to now have been performed for a limited amount of particles ($A<6$) \cite{Li:2019pmg,Lazauskas:2019cxj,Deltuva:2019mnv}. 
This limit is often circumvented by changing the calculation
degrees of freedom and treating the many-body systems as a few-body problem between clusters as done in Ref.~\cite{Arai:2003ek,deDiego:2007rd} introducing approximations in the solution.
Resonances in $^4$H have been studied in the past, for example, in Ref.~\cite{deDiego:2007rd} using an effective two-body n$^3$H potential adjusted to reproduce the p$^3$He experimental phase shifts. 
In this reference, the resonance energy were calculated either directly computing the $S$-matrix pole or by performing a $R$-matrix analysis of the phase shifts, showing a sensible difference among the two methods (the $1^-_I$ resonance is located at energy $E_R=1.2$, $\Gamma=3.5$ MeV, and $E_R=3.6$, $\Gamma=5.3$ MeV, respectively).
This result further enhances the difference between the position of the $S$-matrix pole and the resonance position extracted from the cross section, making difficult any kind of comparison between different calculations and experiments.
Other calculations have been done in Refs.~\cite{Arai:2003ek,Horiuchi:2013iw,Lazauskas:2019cxj,Li:2021ado} using the resonating group method (RGM) with complex scaling, Spin-dipole strength functions, Faddeev-Yakubovsky equations (FYE) in configuration space, and No-core Gamov Shell Model. The results differs from method to method varying in a range $E_R=\{0.9-3.6\}$ MeV and $\Gamma=\{1.0-5.3\}$ MeV.
This variation appears to be a consequence of the different numerical techniques used more than on the interaction employed. In fact, studying the position of the resonance with different realistic interactions but the same theoretical scheme \cite{Lazauskas:2019cxj,Li:2021ado} leads to comparable results.
In this letter, we study the emergence of a 3+1 $J^{\pi}=1^-$ resonance in $^4$H using LO \eftnopi, thus demonstrating the possibility to create n$^3$H $p$-wave states with purely $s$-wave contact interaction. 
In the region of resonance projectile neutron momentum is $\sim 0.5~m_\pi$, binding momenta of nucleons within $^3$H target are $\sim 0.4~m_\pi$ (for a combined distance from the complete disintegration threshold of $0.8~m_\pi$): thus $^4$H resonant states are at the limit of convergence of \eftnopi, nevertheless still observable in a LO calculation.
The absence of any spin-orbit or tensor term in the LO interaction causes degeneracy between $J^{\pi}=0,^-,1^-_I,1^-_{II},2^-$ states which would be splitted including higher orders. 
As $^4$H can still be treated as both a four-body and two-cluster (triton-neutron) system, we choose to employ both the methodologies starting from the same nuclear Hamiltonian.
We begin with a brief description of the interaction theory (LO~\eftnopi) before introducing the
three numerical techniques which were used to reach our main conclusion. The results leading to that
thesis are presented in the third part prior to a summary.

\section{Theory and numerical methods}

The theory used in this work is the \eftnopi\cite{vanKolck:1999mw}~truncated at lowest order (LO).
This LO consists of contact two-body and a contact three-body interactions with corresponding low
energy constants (LECs) fitted to reproduce physical observables. We use the SU(4)
symmetric version of the theory~\cite{Konig:2016utl}. 
In that formulation, nucleon-nucleon spin-singlet and spin-triplet interactions
are identical, and only one contact term appears in the two-body sector. This
increases the uncertainty by $(a_0^{(NN)}m_{\pi})^{-1}\sim30\%$ 
($a_0^{(NN)}$ is the nucleon-nucleon scattering length)
on the top of the~\eftnopi LO truncation error. The contact interactions are 
regularized with a Gaussian cutoff function parametrized by a momentum cutoff $\Lambda$. 
This yields the following two- and three-body potentials in coordinate representation:
\begin{equation}
\hat{V}=C_0 \sum_{i<j} e^{-\frac{r_{ij}^2\Lambda^2}{4}}\quad,
\label{eq:vnn}
\end{equation}
\begin{equation}
\hat{W}=D_0 \sum_{i<j<k} \sum_{cyc}
    e^{-\frac{(r_{ij}^2+r_{ik}^2)\Lambda^2}{4}}\quad.
\label{eq:vnnn}
\end{equation}
The potential receives an identical contribution from each pair of particles if the
relative distance
\footnote{We use boldface characters for three-dimensional vectors, and normal font for
the magnitudes.} $r_{ij}=|\ve{r}_i-\ve{r}_j|$ is small and from every
triplet whose hyperradius - hence the cyclic sum ($cyc$) over each triplet of particle
indices $i<j<k$ - approaches zero for $\lim_{\Lambda\to\infty}$.
 
We fit the cutoff dependence of the $D_0$ LEC to
the triton binding energy ($B_3=8.48$~MeV).
For the two-body LEC $C_0$, we use two different parametrizations in order to assess the
uncertainty of the SU(4) approximation: one at unitarity (with $|a_0^{(NN)}|>10^5$~fm and referred to as ``unitary'') and one fitting the deuteron binding energy $B_2=2.22$~MeV (referred as ``nuclear'').
Due to SU(4) symmetry, all two-nucleon $L=0$ channels exhibit the same proprieties 
implying, \eg, a dineutron bound with the same energy as the triplet $S$-wave neutron-proton system.
We assume a degenerate nucleon mass $m=938.858$ MeV.

Subsequently, we employ three different methods in order to obtain scattering parameters
with this interaction. 
On the one hand, we use two four-body microscopic techniques: Faddeev-Yakubovsky equations (FYE) in configuration space \cite{Lazauskas:2004hq, Lazauskas:2019hil} and the stochastic variational method (SVM)~\cite{Suzuki:1998bn} with harmonic oscillator trap.
On the other hand, we employ a potential-folding technique to obtain an effective two-body problem representing a neutron which
scatters on a solid triton core.

The triton-neutron scattering parameters
($a_L$ and $r_L$) for $L=0$ and $L=1$ are then found trough the effective range expansion (ERE) about the triton-neutron threshold:
\begin{equation}
    k^{2L+1}\text{cot}(\delta_L)=-\frac{1}{a_L}+\frac{1}{2}r_L k^2 + ... \, ,
    \label{eq.app.ere}
\end{equation}
where $\delta_L$ is the triton-neutron phase-shift at angular momentum $L$ and $k$ is the relative momentum of the fragments in the center-of-mass frame. 
Resonances will, therefore, emerge as poles of the system's $T$-matrix:
\begin{equation}
T =\frac{2\pi}{\mu}\frac{1}{\text{cot}(\delta)-i} =
\frac{2\pi}{\mu}\frac{k^{2L+1}}{-\frac{1}{a_L}+\frac{1}{2}r_L k^2 + \dots
-ik^{2L+1}}\, . \label{eq:Tmatrix}
\end{equation}
Here, $\mu \approx 3/4 m$ is the reduce mass of the n$^3$H system.
These poles can be extracted either by fitting phase shifts to the ERE formula or from a direct solution of the Schr\"odinger equation
at complex energies.
We note that the two resonance-extraction methods are not equivalent.

Solving the Schr\"odinger equation for complex energies is affected by the \eftnopi~LO truncation error ($\xi_{\eftnopi}\approx k_{\rm res}/m_\pi\sim80\%$)
, while extracting the resonance position using the ERE of the triton-neutron phase shift is equivalent to the halo-EFT expanded around the n$^3$H threshold.
In the latter case, a new scale, the breakdown momentum of $^3$H, is introduced at $K_{d-n}\simeq94$ MeV/c.
Yet another scale is given by the deuterium-dineutron threshold in the SU(4) case (assuming shallow dimers with $B_2\approx2$~MeV) $K_{d-d}\simeq87\text{MeV/c}$.
Since the scattering parameters are extracted using \eftnopi~at LO, the two methods are expected
to agree up to corrections of the order $\xi_{Halo}\approx k^*_{res}/K_{d-n}\simeq70\%$ 
($\xi_{Halo}\approx k^*_{res}/K_{d-d}\simeq86\%$
for an ERE truncated after the scattering volume.
If the effective range $r_1$ is included, one expects $(\xi_{Halo})^2$.

\vspace{2mm}

\subsection{Faddeev-Yakubovsky-equation formalism}

The Faddeev-Yakubovsky formalism decomposes the few-body Schr\"odinger/Lippman-Schwinger equation into a
set of equations in order to organize the overlapping spectra and divergences of the few-body scattering 
problem. Details about this complex and on our particular numerical realization can be found in
Refs.~\cite{Lazauskas:2004hq,Lazauskas:2019hil}.
The appropriate boundary conditions on the formalism which yield the n$^3$H scattering lengths (volumes)
relevant for this work,
\begin{equation}
\Psi(\tau,\ve{y})=\lim_{k\rightarrow0}\,\left(\psi(\tau)\,\left[j_L(ky)/k^{L}+a_L(k)\, n_L(ky)k^{L+1}\right]\,Y_L(\hat{\ve{y}})\right)
\quad,  \label{eq:FYe_a}
\end{equation}
fix the solutions of the differential FYE equations in the far asymptote where the radial distance $y=\left|\ve{y}\right|$
between the neutron and $^3$H is larger than the support of $\hat{V}$ and $\hat{W}$. The variable $\tau$ in the last equation 
represent set of degrees of freedom of $^3$H system, while $\psi(\tau)$ is the $^3$H ground state wave function whose determination precedes n$^3$H scattering calculations.
The $j_L(z)$ and $n_L(z)$ are spherical Bessel functions of the first and second kind,
respectively, and ERE parameters are extracted in the limit $k\to0~$MeV/c.
In this limit, both terms in the bracket remain finite functions of a distance $|y_{n-^3{\rm H}}|$.

Resonant states are obtained by imposing the canonical boundary conditions on the FYE, namely a purely outgoing wave
in the region where the neutron projectile is no longer interacting with any target $^3$H nucleons:
\begin{equation}
\Psi(t,\ve{y})=\psi(\tau)\,\left[h_L^+(ky)+F_L(k_{\rm res})h_L^-(ky)\right]\,Y_L(\hat{\ve{y}})
\quad.  \label{eq:FYe_res}
\end{equation}
Now, $h_L^-(z)$ and $h_L^+(z)$ are Spherical Bessel functions of the third kind, and 
the solution corresponds to a resonant state for those complex momenta $k$ at which the
function $F_L(k=k_{\rm res})$ vanishes, \ie, where the wave function $\Psi(\tau,\ve{y})$ represents
a purely outgoing n$^3$H waves.
 
\subsection{Harmonic oscillator trap}
The second microscopic approach involves the introduction of a harmonic oscillator (HO) trapping potential
\begin{equation}
    \hat{V}_{\rm HO} = \frac{m}{2A} \omega^2 \sum_{i<j} (\textbf{r}_i - \textbf{r}_j)^2
\end{equation}
with oscillator frequency $\omega$ into $A=4$-body $^4$H system.

Assuming that the range of an interaction is much smaller than the typical HO trap range $R << b_{\rm HO} =\sqrt{2/(m \omega)}$, $\hbar=1$, one can match asymptotic n$^3$H part of a trapped $^4$H wavefunction to the trapped solution of effective n$^3$H two-body system with no nuclear interaction considered. The n$^3$H phasehifts $\delta_L$ at relative momentum $k$ are then extracted using generalized Busch formula for an arbitrary orbital momentum $L$ \cite{Suzuki:2009}    

\begin{equation}
    (-)^{L+1} \left(\sqrt{4\mu \omega}\right)^{2L+1} \frac{\Gamma\left(3/4 + L/2 - \epsilon^n_\omega/2\omega\right)}{\Gamma\left(1/4 - L/2 - \epsilon^n_\omega/2\omega\right)} = k^{2L+1}~{\rm cot}\left(\delta_L \right),
\end{equation}
where $ k=\sqrt{2\mu \epsilon^n_\omega}$, $\Gamma(x)=\int_0^\infty z^{x-1}e^{-z}dz$, and $\epsilon^n_\omega = E_\omega^n (^4{\rm H})-E_\omega (^3{\rm H})$ is an energy of the $n$-th $^4$H excited state in a trap with respect to the n$^3$H threshold. Here, bound-state energies $E_\omega (^3{\rm H})$, $E_\omega^n (^4{\rm H})$ are calculated using Stochastic Variational Method \cite{Suzuki:1998bn}. Both $s$-wave ($L=0$) and $p$-wave ($L=1$) n$^3$H phase shifts are extracted applying HO trap lengths $40~{\rm fm} \leq b_{\rm HO} \leq 60$~fm.        

\begin{figure*}
\centering
  \centering
  \includegraphics[width=\linewidth]{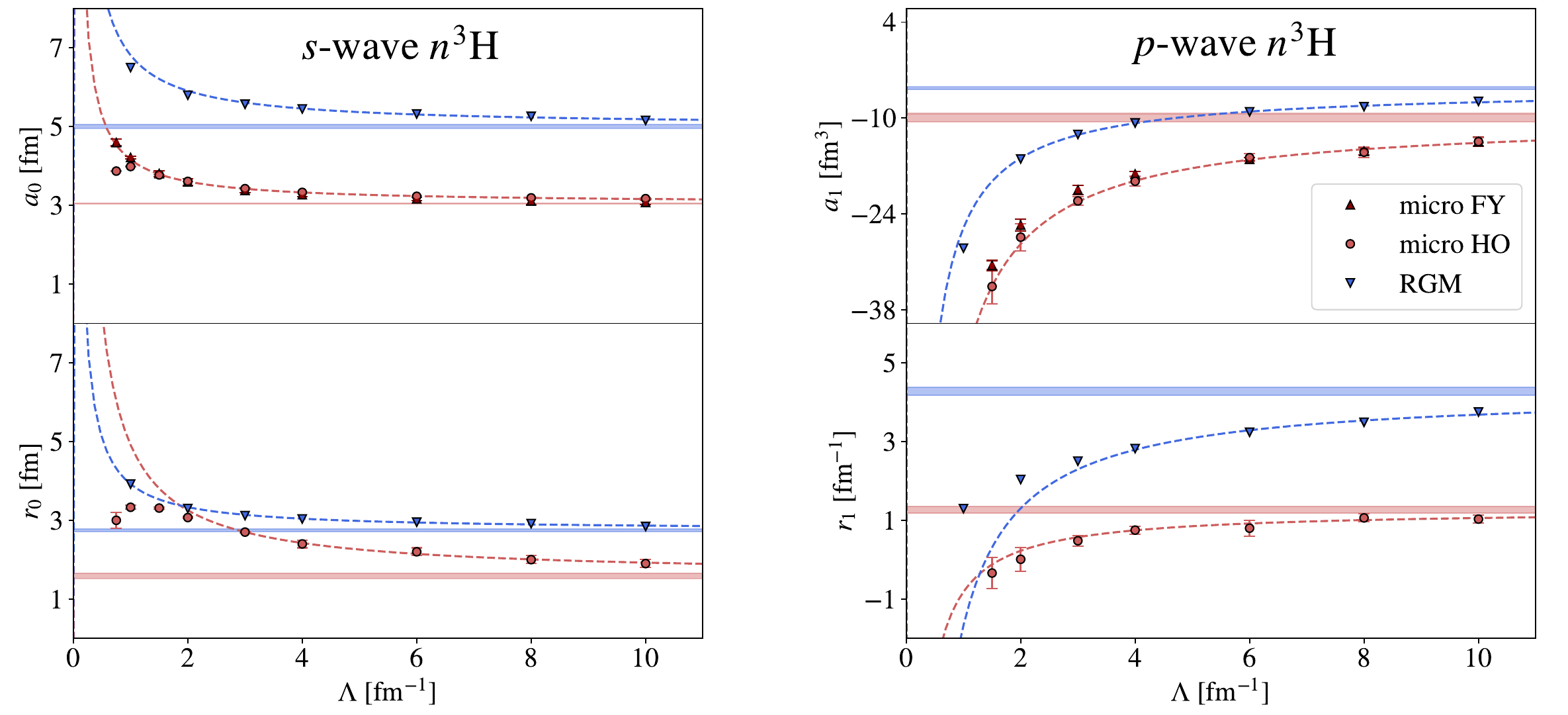}
  \caption{Left panel: $s$-wave scattering length $a_0$ and effective range $r_0$ calculated using microscopic techniques, SVM with Busch Formula (red circles and triangles) and solving the Faddeev-Yakubovsky equation (red triangles), and n$^3$H folding using RGM (blue) for the "nuclear" case. Right panel : the same as in the left panel but for $p$-wave scattering parameters - scattering volume $a_1$ and $p$-wave effective range $r_1$. For each scattering parameter the corresponding dashed line represents an extrapolation to the contact limit ($\Lambda \rightarrow \infty$) using the function in~\eref{extrapolation}. The extrapolated values together with extrapolation uncertainties are shown by the shaded bands.}
  \label{fig:ERE_Parameters}
\end{figure*}

\subsection{Resonating Group Method}
For neutron momenta much smaller than any excitation
scale of the triton, $k\ll K_{d-n}$, one expects contributions to the scattering
event from distortions of projectile and target to be negligible.
The resonating-group idea (pedagogical:~\cite{wildermuth1977unified}, original:~\cite{Wheeler:1937zz})
realizes this expectation with the lowest order of
its expansion of the wave function of the four-body problem.
In general, this expansion is in an overcomplete basis of
products of fragment-internal ($\phi$) and relative ($\chi$) states: 
\begin{equation}\label{eq:rgm_wfkt}
\Psi=\sum_j^{|\alpha|}\prod_{m=1}^{|\alpha_j|}\phi^j_m\prod_{n=1}^{|\alpha_j|-1}\chi^j_n
\quad.
\end{equation}
The cluster states $\phi^j$ resemble partitions ($\alpha_j$) of the $A$ nucleons
into bound states which appear at the energies of interest.
We employ the same boundary conditions as in the FY approach~\eref{eq:FYe_a} but we
look for solutions of the four-body Schr\"odinger equation in a subspace in which
only variations of the relative function are admitted. 
Thereby, the four-body problem reduces to an effective one-body problem for
the relative motion between the triton and neutron.
In the notation of \eref{eq:rgm_wfkt}, a single partition,
$\alpha=\lbrace(\text{triton, neutron})\rbrace$ comprising
two fragments $\phi_1^1=\psi_t$ and $\phi_2^1=\psi_n=1$ is used.
With a Gaussian expansion of the extended three-body core $\psi_t$, the Gaussian regulated
EFT potentials translate analytically into an effective
core-atom interaction with local and non-local, energy-dependent 
components. This effective potential takes the
permutation properties of the neutron projectile and its core copy fully into account:
\begin{equation}
\left(\hat{T}_{\rm rel}-E+\hat{U}^{(2)}_{\rm eff}(\ve{y})\right)\chi(\ve{y})+
\int\hat{U}^{(3)}_{\rm eff}(\ve{y},\ve{y}',E)\chi(\ve{y}')d\ve{y}'=0
\quad.
\label{eq:rgm_sgl}
\end{equation}
We then solve this integro-differential equation, 
which is projected into partial waves, to obtain the relative $s$- and $p$-wave functions via a standard finite-difference method. The amplitude 
$a_L(k)$ is subsequently found with boundary conditions identical to~\eref{eq:FYe_a},
albeit, for real momenta.
It is worth noting that this structure can be interpreted as the dynamical equation of a trimer-atom halo EFT (review:~\cite{Hammer:2022lhx}). 
This relation between the RGM, extended-core treatment and a field theory with elementary neutron and trimer point-like fields becomes explicit through a local expansion~\cite{Apagyi_1983}~of the integral kernel in \eref{eq:rgm_sgl}.
The non-local part of \eref{eq:rgm_sgl}~is thereby expressed as a series 
\begin{equation}\label{eq:rgm_nonlocexp}
\int\hat{U}^{(3)}_{\rm eff}(\ve{y},\ve{y}',E)\chi(\ve{y}')d\ve{y}'
\to
\left[
\sum_{n=0}^\infty\hat{K}_n(\ve{r},E)\ve{\nabla}^{(n)}_{\ve{r}}
\right]
\chi(\ve{r})
\end{equation}
whose summands define the vertices of the infinite terms of a Lagrangean built of
trimer and atom fields. In principle, our choices of regulators and core-wave-function expansion
allow for an analytic evaluation of this vertex structure. This is beyond the
scope of this work where it suffices to realize the general structure of the $\hat{K}_n$ operators
as products of order-$n$ polynomials times Gaussians. The strength and fall-off parameters of each
$\hat{K}_n$ are, in principle, determined through the atom-atom renormalization conditions implicit
in the LECs $C_0$ and $D_0$.

Stated explicitly, no additional renormalization of the effective triton-neutron potentials
\footnote{Written in the so-called parameter representation,
$\ve{y}'$ is an additional expansion label. As such, permutation operators
have no effect on it. Hence, one does not expect a symmetric form
for the kernel $\hat{U}^{(3)}_{\rm eff}(\ve{y},\ve{y}',E)$ in which
the core-projectile exchange has been already considered
(see~\figref{Fig:Potential}).}~
is required, and we obtain their dependence
on $C_0$ and $D_0$~(see~\eref{eq:vnn,eq:vnnn})~by analytically 
integrating out fast, \ie, core-internal degrees of freedom.
The LECs of the halo EFT, \ie, the terms in \eref{eq:rgm_nonlocexp}~relevant
at a certain order, inherit their dependence from $\hat{U}_{\rm eff}$. 
In effect,~\eref{eq:rgm_sgl}~represents a halo EFT with two degrees of freedom:
the extended but inert triton
and a neutron, with an interaction parametrized by the underlying nuclear theory.
A necessary matching constraint between the two descriptions 
is the wave function of the three-nucleon state, \ie, the core.
To satisfy this constraint, we feed the single-particle 
densities of the SVM calculation
to the RGM integration over the fast degrees of freedom, thereby obtaining $\hat{U}^{(2)/(3)}_{\rm eff}$
as functions of the expansion coefficients of these densities in a set of
$\leq8$ Gaussian functions.
An important difference to the microscopic formulations is the
smaller breakdown momentum $K_{d-n}$.
Compared to $K_{d-n}$, the resonance momentum remains
relatively large even when calculated with
respect to the n$^3$H threshold ($k_{\rm res}^{halo}$),
as pertinent for halo-EFT formulation.
Therefore, the na\"ive estimate for the theoretical
uncertainty of the halo EFT is larger than the
expected~\eftnopi~uncertainty: $k_{\rm res}^{halo}/K_{d-n}>k_{\rm res}/m_\pi$.
However, the renormalizability of the theory,~\ie, the cutoff
independence inside the truncation error of the theory for
the resonance-related observables, and the sheer existence of
the pole are accessible with the RGM formulation.

\section{Results and discussion}

\begin{figure*}
\centering
\includegraphics[width=\textwidth]{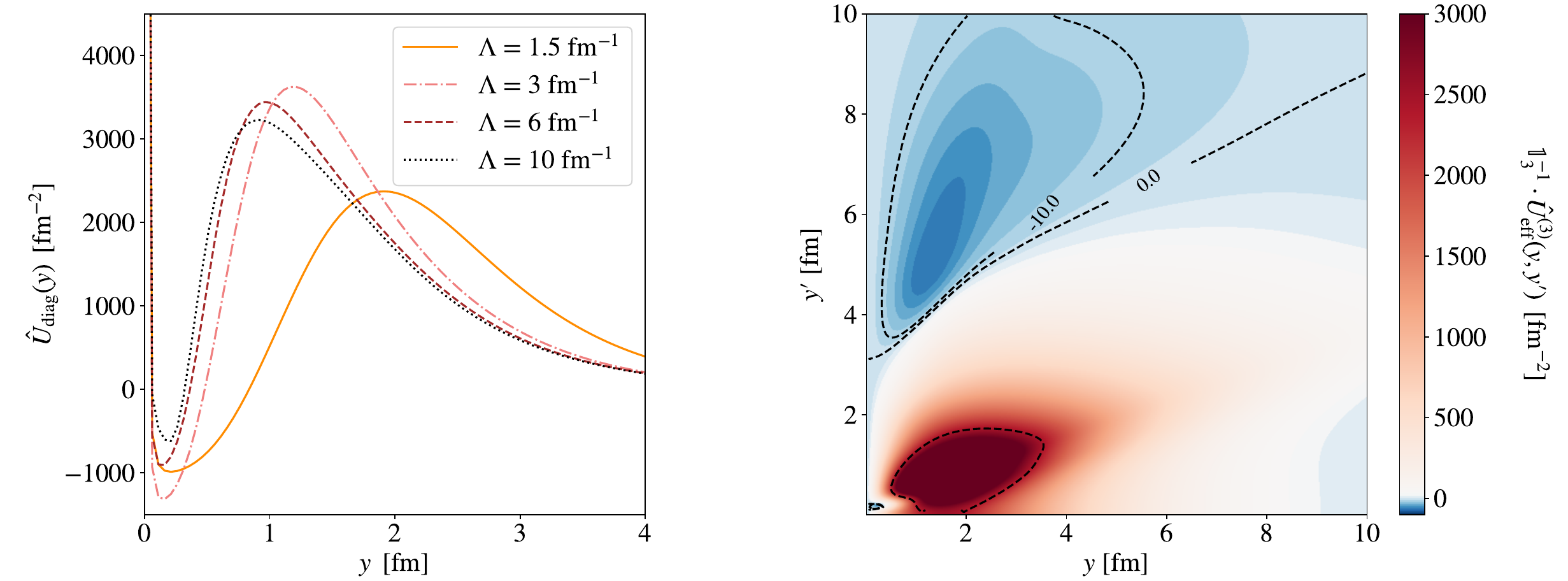}
\caption{Left panel:
Diagonal part of the effective ``nuclear''  RGM potential in
the $L=1$ partial wave, 
\mbox{$\hat{U}_{\rm diag}(y):=
L(L+1)y^{-2}+
\frac{2\mu}{\hbar^2}\left[
\hat{U}_{\rm eff}^{(2)}(y)+
\hat{U}_{\rm eff}^{(3)}(y,y)y^2
\right]\frac{1\text{fm}}{N_3} $},
where $N_3$ is the norm of the triton wave function
and
$1.5\,{\rm fm}^{-1}\leq\Lambda\leq10\,{\rm fm}^{-1}$. 
Right panel:
Topography of the non-local part $\hat{U}_{\rm eff}^{(3)}(y,y')$ ($\Lambda=10$~fm$^{-1}$)
 exhibiting an off-diagonal attractive pocket (second quadrant of the contour plot) which does not vanishes in the zero-range limit.  
 The potentials were obtained at $E=0$ with a five-dimensional Gaussian calibration of the core wave function
 to SVM ground states.}
 \label{Fig:Potential}
\end{figure*}

We begin discussing \eftnopi LO predictions in the n$^3$H scattering system with our
results for the effective-range parameters $a_{L}$ and $r_{L}$.
Their cutoff-$\Lambda$ dependence is shown in~\figref{fig:ERE_Parameters},
where we compare values obtained with the four-body formulations (red circles) with
the RGM extractions.
We find the difference between the former two, \ie, HO trap and FYE, $\simeq 0.1-1 \%$ for $a_0$
and $\simeq 1-10\%$ for $a_1$. The numerical uncertainty for these two approaches is depicted by the errorbars.

The observed behavior for all parameters
resembles the LO-\eftnopi-characteristic $1/\Lambda$ falloff. 
For $\Lambda \geq 4~{\rm fm^{-1}}$, we thus fit the function
\begin{equation}
    O(\Lambda) = O(\infty) +\frac{\alpha}{\Lambda}
    \label{extrapolation}
\end{equation}
in order to extrapolate 
the
contact/zero-range limit ($\Lambda \rightarrow \infty$). 
We find
$a_0(\infty) = 3.05(1)$~fm, $r_0(\infty) = 1.59(7)$~fm and $a_1(\infty) = -9.9(6)$~fm$^3$, $r_1(\infty) = 1.3(1)$~fm$^{-1}$,
with an uncertainty set by extrapolation errors.
For the ``unitary'' $NN$ case, $a_0(\infty) = 2.27(1)$~fm, $r_0(\infty) = 0.52(3)$~fm, and
$a_1(\infty) = -5.1(1)$~fm$^3$, $r_1(\infty) = 1.8(1)$~fm$^{-1}$. 
With the sign convention of~\eref{eq.app.ere}, $a_0>0$ relates to an overall repulsive $s$-wave interaction,
as intuitively expected for Pauli forbidden states. 
Consistently, we find no bound-state solutions in the $L=0^+$ n$^3$H channel for any of the considered $\Lambda$s.
In contrast, $a_1<0$ for all $\Lambda$ indicates the presence of attractive $p$-wave n$^3$H interaction.
With decreasing cutoff, $|a_1|$, and the strength of the interaction, increases
until a bound state emerges at $\Lambda<0.75\,{\rm fm}^{-1}$. 
%

%
The character of an attractive odd partial wave and a repulsive even partial wave interaction
is found in the nuclear and unitary scenarios. 
Scattering length and volume
decrease when removing the two-body scale in the unitary limit.
Compared with a three-body scale set by $R_3=1/\sqrt{-2mE_3/\hbar^2}\simeq 6$~fm, \ie, the only scale present in the
unitary system, the scattering parameters are natural: $|a_0(\infty)|\simeq0.4\,R_3$, $|r_0(\infty)| \simeq 0.1\,R_3$, $|a_1(\infty)|^{1/3} \simeq 0.3\,R_3$, and $|r_1(\infty)|^{-1} \simeq 0.1\,R_3$. 
%

%
Our four-body and RGM results for $a_0$ match sign and magnitude of the calculations and experiments available in literature (within~\eftnopi~uncertainty). 
This qualitative agreement holds in the ``nuclear'' and the ``unitary'' cases. 
Specifically, we compare our results to the coherent scattering length $a_0^c=\nicefrac{1}{4} a^{S=0}_0 + \nicefrac{3}{4} a^{S=1}_0$, and
use $a_0^c({\rm theo}) \approx 3.6$ fm~\cite{Lazauskas:2019hil}~and $a_0^c({\rm exp})=3.5\,-\,3.9$ fm~\cite{Tilley:1992zz}~as theoretical and experimental benchmarks.

\begin{figure*}
\centering
  \includegraphics[width=.98\linewidth]{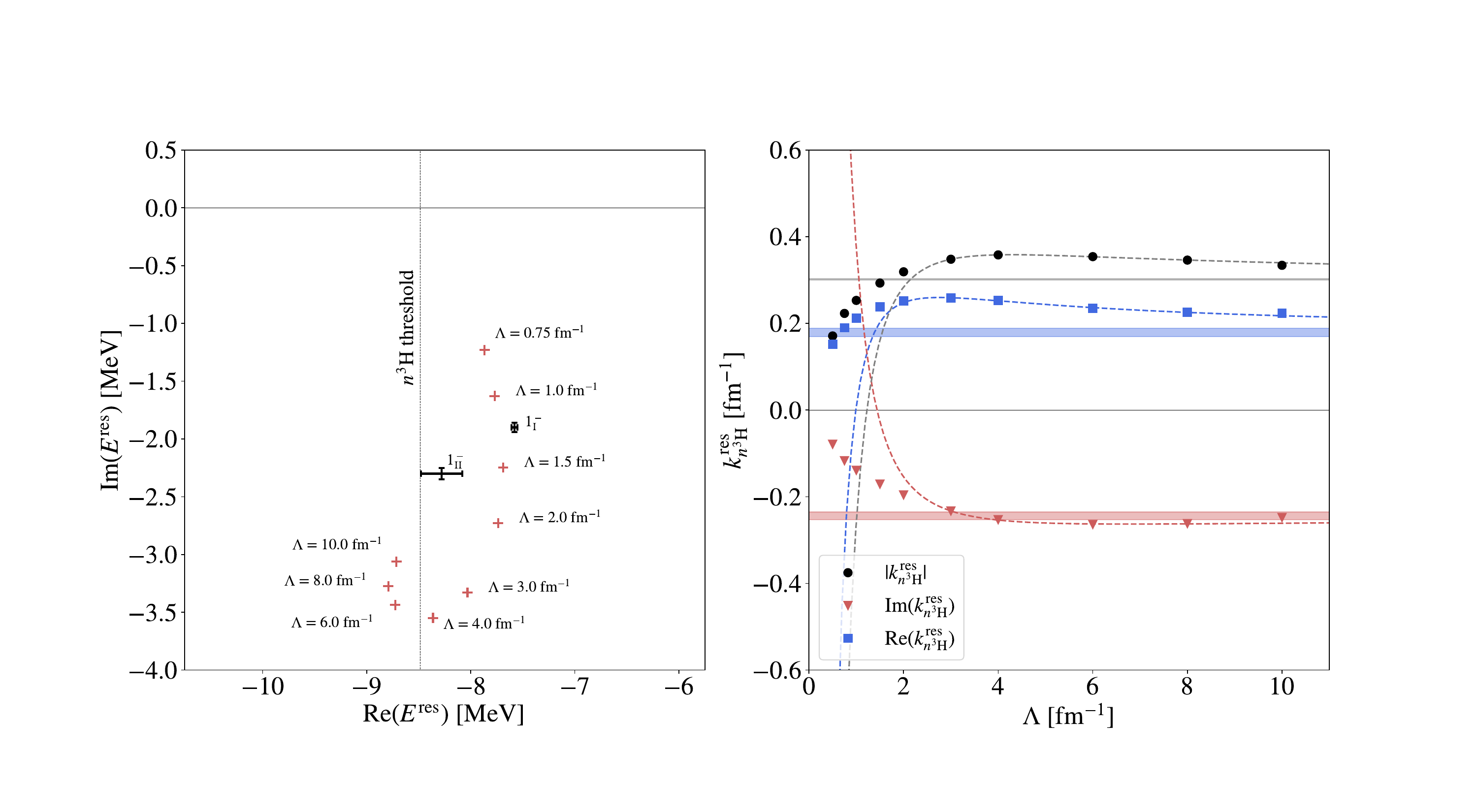}
  \caption{ Left panel : Complex energy plane with calculated $^4$H $1^-$ resonance positions (red crosses) for multiple momentum cutoff $\Lambda$. 
  The black error bars represent the position of the $1_I^-$ and $1_{II}^-$ resonances calculated in Ref~\cite{Lazauskas:2019cxj} for realistic interaction model.
  Right panel: The absolute value, real, and imaginary part of the complex resonance momentum $k_{res}$ with respect to the $n^3$H threshold as a function of increasing $\Lambda$. The momentum is related to resonance energies via~\eref{mmnt}. The resonance pole was extracted directly solving the four-nucleon problem trough the Faddeev-Yakubovsky equations. Dashed lines and the colored bands represent the extrapolation of the results done using a three parameter fit (up to $\left(1/\Lambda\right)^2$) using cutoffs $\Lambda>3~{\rm fm}^{-1}$}
  \label{fig:Pole_PositionN}
\end{figure*}

The RGM predictions for the ``nuclear'' parametrization are
$a_0(\infty) = 5.71$~fm, $r_0(\infty) \simeq 3.15$~fm, $a_1(\infty) = -10.49$~fm$^3$, $r_1(\infty) = 4.33$~fm$^{-1}$, and
$a_0(\infty) = 3.3$~fm, $r_0(\infty) \simeq 1.8$~fm, $a_1(\infty) = -2.5$~fm$^3$, $r_1(\infty) = 10.1$~fm$^{-1}$ for the ``unitary'' case.
These findings are consistent with the above in sign and thus with the attractive/repulsive character but differ in absolute value, providing a more repulsive interaction. 
This is consistent with the lack of excitations in the considered triton core, which naively introduces a variational shift in the system groundstate.
Still RGM results retain the same functional dependence on $\Lambda$ for all scattering parameters.
Regarding renormalization-group dependence, in particular, on the stability of pole varying the cutoff, both theories are apparently equally useful for the amplitudes under consideration.

Although the effective n$^3$H interaction is not an observable, we use the RGM to extract its $p$-wave
component to acquire a deeper insight into the interplay between repulsive and attractive parts and their evolution with $\Lambda$.
This allows for a heuristic assessment of whether or not one should expect a resonant-type solution.
In~\figref{Fig:Potential}, we visualize this interaction for the "nuclear" case in the zero-energy approximation.
In the left panel, we display the dependence of the sum of the angular momentum barrier, the local and
non-local (at $y'=y$ ) parts
of the coordinate space interaction,~\eref{eq:rgm_sgl}, as a function of the relative coordinate $y$ between the triton
center of mass and the neutron.
For small cutoffs, the potentials form a positive energy pocket at $y\simeq0.5$~fm.
The pocket vanishes in the contact limit $\Lambda\to\infty$, where the potential is
visibly incapable of forming quasi-stationary states like resonances.
While the pocket suggests the possibility to form resonant states at small
cutoffs, we find that its formation demands the inclusion of the non-local
interaction, \ie, one must take into account the exchange between projectile and core neutron. 
In the right panel, we investigate the character of the non-local part with a contour plot
at $\Lambda=10$~fm$^{-1}$ and zero energy.
The dominance of a repulsive part (red) prevails in the non-local potential for most points 
with the notable exception of a weak attractive region $min(\hat{U}_{\rm eff}^{(3)})\approx$-0.35 MeV; blue) far off the diagonal where $y'>y$. 
The pocket, which does not vanish in the zero contact limit, is apparently too shallow to support a bound state, but it might suffice to form a broad resonance state.

%
With the Faddeev-Yakubovsky approach, we search for a pole corresponding to a resonance in the $J^\pi=1^-$ channel, directly.
The results of this search are displayed in~\figref{fig:Pole_PositionN}.
In the left panel, the cutoff dependence of the pole position in the complex energy plane is shown (``nuclear'' case). 
The trajectory passes the n$^3$H threshold, $B_3=8.484$~MeV (dashed line) and stabilizes for $\Lambda\gtrsim6~{\rm fm}^{-1}$
at $E_{\rm res}=-8.7-i3.1$ MeV. 
In order to study the pole convergence, we translate resonance energies to complex n$^3$H momenta $k_{\rm res}$:

\begin{equation}
    k_{\rm res} = \sqrt{2 \mu \tilde{E}_{\rm res}},~~~~\tilde{E}_{\rm res}= E_{\rm res} - E_{t}
    \quad.
    \label{mmnt}
\end{equation}

The $\Lambda$ dependence of ${\rm Re}(k_{\rm res})$, ${\rm Im}(k_{\rm res})$, and $|k_{\rm res}|$ is shown
in the right panel of~\figref{fig:Pole_PositionN}.  
The resonance momentum stabilizes for $\Lambda\gtrsim m_\pi$, \ie, the breakdown scale of the theory.
Its absolute value $|k_{\rm res}|$ does not exceed $0
.35~{\rm fm}^{-1}\simeq 0.5~m_\pi$.
The only point that seems to deviate from the the converged behaviour is $\Lambda=10~{\rm fm}^{-1}$, however, this also correspond to the most unstable calculations where we expect the largest numerical uncertainties.

Alternatively, the resonance position can also be found as pole of the effective-range-expanded n$^3$H $T$-matrix in~\eref{eq:Tmatrix}. 
With the ERE parameters, $a_1$
, calculated with four-body methods, we find $k_{\rm res}^{\rm halo}\simeq0.41-0.23\,i~{\rm fm}^{-1}$ ( $E_{\rm res}^{\rm halo} = 3.2 -i\, 5.2 i$ MeV. 
The value is close to the estimated breakdown $K_{d-n}$ of the halo approach implying a
relatively large uncertainty.
Within this uncertainty margin, $k_{\rm res}^{\rm halo}$ is consistent with the
pole obtained with the FY method: 
$k_{\rm res}\simeq0.23-0.25\,i~{\rm fm}^{-1}$ for $\Lambda=10$ fm$^{-1}$.
For reasons stated in the introduction regarding the model-dependence
in data, and the ensuing
discrepancies in experimental values, we abstain
from a comparison of our predictions to data.
To justify this with another example, we note
that the real part of the energy of a $S$-matrix pole is,
in general, found smaller than the one extracted with a $R$-matrix analysis.
This behavior is exemplified by the  transfer reactions $^2$H(t,p)$^4$H and $^3$H(t,d)$^4$H~\cite{Sidorchuk:2004ntw}. 
The $R$-matrix analysis of the results obtains for the lowest $^4$H state
$E_R=3.05\pm0.19$~MeV and $\Gamma=4.18\pm1.02$~MeV. The corresponding $S$-matrix pole,
extracted by the same authors,
led to $E_R=1.99\pm0.37$~MeV and $\Gamma=2.85\pm0.3$~MeV. 
Therefore, a fair comparison can only be made with other theoretical $S$-matrix-pole predictions. 
For a quantitative comparison,
we consider the pole location predicted in~\cite{Lazauskas:2019cxj,Lazauskas:2019hil,Arai:2003ek, deDiego:2007rd}:
$E_R=(0.9\,-\,1.23)$~MeV and $\Gamma=(3.5\,-\,5.8)$~MeV.
The~LO~\eftnopi pole is broader and subthreshold: $E_R=-0.22$~MeV and $\Gamma=6.2$~MeV.
The size of the deviation appears to be consistent with the effect induced by SU(4) symmetry breaking found in~\cite{Lazauskas:2019cxj} (\ie, the difference between two black error bars in~\figref{fig:Pole_PositionN}).
On the point of view of~\eftnopi, the deviation
is also consistent with the expected truncation uncertainty.
Finally, for the ``unitary'' parametrization, the FY calculations find a less pronounced resonance
compared with the ``nuclear'' case.
We find the numerical tracing with $\Lambda$ prohibitive and practical only for cutoffs of the order of $m_\pi$. 
We are not able to determine exact position of the $S$-matrix pole in the zero-range limit $\Lambda\to\infty$, as this pole resides far away from the real momenta axis and leads to a strong divergence of the outgoing waves, making FY numerical procedure unstable.
For $\Lambda=2$ fm$^{-1}$ the resonance has an energy $E_R=-8.156$~MeV and $\Gamma=9.6$~MeV (corresponding to an absolute $k_{res}=0.58\,m_\pi$). 
Notice that since the differences between the SU(4) ``nuclear'' and ``unitary'', as well as the the deviation induced by a broken SU(4), are all subleading and perturbative: the difference between the results are expected to deviate only by a shift in the state positions.
In other world, the presence of a resonance for a finite cutoff in the ``unitary'' case and the convergence of the ``nuclear'' data imply also the convergence of the ``unitary'' pole.
Moreover, the same reasoning would apply to the case in which the SU(4) symmetry is broken already at LO of the theory.
However, the impossibility of estimating the converged value of the position of the ``unitary'' resonance pole does not allow to set a clear area where the non-SU(4) result will fall.
Nonetheless, the ``unitary'' pole is still expected to be inside the truncation error of the theory and the deviation with respect to the ``nuclear'' one is in line with the expected $\approx30\%$ error for the SU(4) theory truncation.
Our findings for total isospin $T$=1 system correlate with ones of ref. \cite{Deltuva:2011djw} obtained for $T$=0 system, in which the authors 
also find diminution of the resonance effect on scattering data once  cutoff increases. 
In fact, once  for larger cutoffs the pole moves subtreshold its effect on the calculated phaseshifts 
becomes hardly visible.

\section{Conclusions}
%
In this work, we predict, model-independently, the existence of a cutoff-stable $J^{\pi}=1^-$ resonance in $^4$H
with the zero-range theory of nuclear forces: leading-order pionless effective field theory (\eftnopi).
We reach this result with three independent numerical techniques, two microscopic methods, and an
analytically-matched halo effective theory.
This finding of a cutoff-stable $p$-wave resonance complements and generalizes studies on such states with pure contact theories
which were, as of now, limited to $s$-wave channels. 
%
Our results agree qualitatively with each other and with earlier simulations which address the existence of
a $^4$H resonance with high-precision interaction models.

Most remarkable is the emergence of $p$-wave poles from a $s$-wave contact theory between two and three particles. 
The major problem of the theory in many-nucleon systems is the absence of stable bound states for nuclei larger than $^4$He at LO.
The possibility of finding resonances in $p$-wave systems opens the chance that the ground states of larger nuclei appear as resonances. 
If this is the case, we conjecture such poles may be then moved on the stable region by the perturbative insertion of sub-leading orders of the theory.
This transition of $p$-wave resonances into bound states, obviously, cannot occur in the absence of the former.
More studies will be needed in the future to check this eventuality.

\section*{acknowledgement}
We thank H.~W.~Grie\ss hammer, N.~Walet, and U.~van Kolck for helpful discussions.
MS was supported by the Pazy Foundation and the Israel Science Foundation grant 1086/21 and
JK by STFC~ST/P004423/1~and the US Department of Energy
under contract DE-SC0015393.
This research was completed during the program Living Near Unitarity at the Kavli Institute for Theoretical Physics (KITP),  University of Santa Barbara (California),
and is supported in part by the National Science Foundation under Grant No. NSF PHY-1748958.
We were granted access to the HPC resources of TGCC/IDRIS under the
allocation 2022-A0110506006 made by GENCI (Grand Equipement
National de Calcul Intensif).

\bibliographystyle{ieeetr}

\bibliography{PLB.bib}

\end{document}